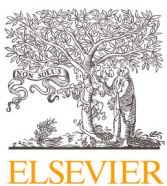
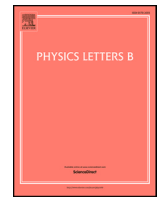

Letter

# Grand unified origin of enhanced scalar couplings: Connecting radiatively broken electroweak symmetry to SO(10) dynamics

Farrukh A. Chishtie  a,b

[a] *Peaceful Society, Science and Innovation Foundation, Vancouver, BC, Canada*
[b] *Department of Occupational Science and Occupational Therapy, University of British Columbia, Vancouver, BC, Canada*

A B S T R A C T

We propose that the enhanced Higgs quartic coupling required by radiatively broken electroweak symmetry (RBEWS) emerges naturally from SO(10) grand unification. Our previous analysis demonstrated that a coupling enhancement factor $k = \lambda_{\text{enhanced}}/\lambda_{\text{SM}}$ leads to absolute vacuum stability with a UV Landau pole near the GUT scale for $k \gtrsim 1.03$. The RBEWS prediction $e_{125} = 7.2$ of Steele and Wang, when properly translated from the Coleman-Weinberg scheme at the electroweak VEV to the $\overline{\text{MS}}$ scheme at $M_t$ via scheme conversion and scale-dependent ratio evolution, yields $k(M_t) \approx 6.0$–$6.4$, corresponding to a UV pole at $\Lambda_{\text{UV}} \sim 1.5$–$2 \times 10^{16}$ GeV—remarkably close to the GUT scale $M_{\text{GUT}} \sim 2 \times 10^{16}$ GeV. We argue this coincidence is not accidental: the UV pole signals the scale where the Standard Model effective description must be embedded into the full SO(10) structure. We derive threshold corrections from SO(10) scalar sectors containing $\mathbf{10}_H$, $\overline{\mathbf{126}}_H$, and $\mathbf{45}_H$ representations, showing that portal couplings between the light Higgs doublet and heavy GUT scalars can generate enhancement factors of order $k \sim 5$–$10$ at the matching scale. The Coleman-Weinberg mechanism operating within a classically scale-invariant GUT scalar potential provides a dynamical origin for both RBEWS and the hierarchy between $M_{\text{GUT}}$ and the electroweak scale. The properly translated enhancement $k(M_t) \approx 6.0$–$6.4$ predicts a trilinear coupling modifier $\kappa_\lambda$ consistent with the current ATLAS constraint $\kappa_\lambda < 6.6$ at 95% CL from combined Run 2+3 $HH \to b\bar{b}\gamma\gamma$ data, while offering a definitive test at the HL-LHC. Our framework further predicts correlations with proton decay rates and gravitational wave signatures from cosmic strings, providing multiple experimental probes of the unified scenario.

## 1. Introduction

The discovery of the 125 GeV Higgs boson [1,2] confirmed the mechanism of electroweak symmetry breaking while simultaneously exposing a theoretical puzzle: the Standard Model (SM) Higgs quartic coupling $\lambda(M_t) \approx 0.126$ [3] runs negative at scales $\mu \sim 10^{10}$ GeV, rendering the electroweak vacuum metastable [4]. This precarious proximity to the stability boundary suggests deeper physics.

In a recent work [5], we demonstrated that SM vacuum stability exhibits extreme sensitivity to the Higgs quartic coupling. A critical enhancement factor $k_{\text{crit}} = 1.03$ separates metastability from absolute stability with UV Landau poles. In the radiatively broken electroweak symmetry (RBEWS) scenario of Steele and Wang [6], which predicts $k \approx 7.2$, places the theory deep in the absolutely stable regime with a UV pole at

$$\Lambda_{\text{UV}} \approx 1.5 \times 10^{16} \text{ GeV}, \tag{1}$$

remarkably close to the grand unified theory (GUT) scale $M_{\text{GUT}} \sim 2 \times 10^{16}$ GeV where gauge couplings unify [7,8].

This proximity suggests a profound connection: the enhanced scalar coupling responsible for RBEWS may originate from GUT-scale dynamics. Rather than viewing the UV pole as pathological, we interpret it as signaling the scale where the SM effective field theory (EFT) description breaks down and must be embedded into a more complete structure—precisely the SO(10) grand unified theory [8,9].

The theoretical foundations of RBEWS have been developed extensively through renormalization group methods for effective potentials. Early work established the all-orders summation of leading-logarithm contributions to radiatively broken electroweak symmetry in the electroweak sector [10,11], revealing that while the top-quark Yukawa coupling poses challenges for small quartic coupling values, large $\lambda$ scenarios remain viable. Subsequent analyses extended these methods to subleading logarithm orders [12,13], demonstrating the stability of radiative corrections and establishing systematic techniques for computing effective potentials to high loop orders [14]. These renormalization group summation methods have proven essential for understanding non-perturbative aspects of symmetry breaking and provide the technical foundation for the RBEWS mechanism we connect to GUT physics in this work.

In this paper, we develop the scenario where RBEWS emerges as a consequence of SO(10) symmetry breaking rather than an independent phenomenon. We show that threshold corrections from SO(10) scalar representations naturally generate $\mathcal{O}(1)$ enhancements of the Higgs quartic coupling at the GUT scale. Portal couplings between the light Higgs doublet and heavy GUT scalars provide the microscopic origin for enhancement factors $k \sim 5$–$10$. The Coleman-Weinberg mechanism [15] operating within the multi-scalar GUT potential can dynamically generate both RBEWS at low energies and the GUT symmetry breaking pattern. Specific breaking chains through intermediate symmetries (Pati-Salam or left-right) constrain the viable parameter space.

Our framework connects three seemingly disparate phenomena: the 125 GeV Higgs mass, vacuum stability, and grand unification. This synthesis offers multiple experimental tests through precision Higgs self-coupling measurements at future colliders, proton decay searches at Hyper-Kamiokande [16], and gravitational wave signatures from cosmic strings [17].

The paper is organized as follows. Section 2 reviews the sensitivity analysis and RBEWS predictions. Section 3 develops the SO(10) framework and threshold corrections. Section 4 analyzes portal couplings and enhancement mechanisms. Section 5 discusses the Coleman-Weinberg mechanism in the GUT context. Section 6 examines phenomenological implications. We conclude in Section 7.

## 2. Review: Vacuum stability sensitivity and RBEWS

### 2.1. Enhancement factor parameterization

Following our previous analysis [5], we parameterize enhanced scalar coupling scenarios through

$$k = \frac{\lambda_{\text{enhanced}}(M_t)}{\lambda_{\text{SM}}(M_t)}, \tag{2}$$

where $\lambda_{\text{SM}}(M_t) = 0.12604 \pm 0.00030$ [3]. This isolates the physics of enhancement from technical scheme-dependence issues.

The three-loop beta function for the Higgs quartic coupling [3,18] takes the schematic form

$$\beta_\lambda = \frac{1}{16\pi^2}\left[12\lambda^2 + 6\lambda y_t^2 - 3y_t^4 + \cdots\right] + \beta_\lambda^{(2)} + \beta_\lambda^{(3)}, \tag{3}$$

where the crucial competition is between the positive $12\lambda^2$ self-interaction term and the negative $-3y_t^4$ top Yukawa contribution.

### 2.2. Phase structure

Our systematic analysis revealed three distinct phases as function of $k$. For $k < 1.03$ (Region I, Metastable), the coupling turns negative at an instability scale $\Lambda_I$, with $\Lambda_I \sim 10^{10}$ GeV for the SM ($k = 1$). At $k_{\text{crit}} = 1.03$ (Region II, Marginal), the coupling asymptotes toward $\lambda \to 0^+$ without crossing zero—the phase boundary. For $k > 1.03$ (Region III, Absolutely Stable with UV Poles), the vacuum is absolutely stable but the positive $\lambda^2$ feedback drives runaway growth toward a Landau pole at $\Lambda_{\text{UV}}$.

The RBEWS scenario with $k \approx 7.2$ corresponds to Steele and Wang's result [6]:

$$e_{125} = \frac{\lambda_{\text{RBEWS}}(v)}{\lambda_{\text{CSB}}(v)} = \frac{0.23}{0.032} \approx 7.2, \tag{4}$$

where the radiatively generated coupling is enhanced relative to conventional symmetry breaking at the electroweak scale.

Here $\lambda_{\text{CSB}}$ denotes the conventional symmetry breaking (CSB) quartic coupling in the Coleman-Weinberg scheme of Steele and Wang [6], evaluated in the O(4) scalar field normalization $V = \lambda \phi^4$. The CSB value $\lambda_{\text{CSB}}(v) = 0.032$ corresponds to $4\lambda_{\text{CSB}} = 0.128 \approx \lambda_{\text{SM}}(M_t)$ in the $\overline{\text{MS}}$ scheme with SU(2) doublet normalization $V = \lambda|H|^4$ [3].

### 2.3. Coupling conventions and scheme dependence

The enhancement factor $e_{125} = 7.2$ of Ref. [6] is defined as the ratio of Coleman-Weinberg (CW) scheme quartic couplings at the electroweak VEV $v = 246$ GeV:

$$e_{125} = \frac{\lambda_{\text{RBEWS}}^{\text{CW}}(v)}{\lambda_{\text{CSB}}^{\text{CW}}(v)} = \frac{0.23}{0.032} \approx 7.2, \tag{5}$$

where $\lambda_{\text{CSB}}^{\text{CW}}$ denotes the conventional symmetry breaking coupling in the O(4) scalar field normalization $V = \lambda \phi^4$ of Steele and Wang. The SM value $\lambda_{\text{SM}}(M_t) = 0.12604$ [3] uses the $\overline{\text{MS}}$ scheme with the SU(2) doublet normalization $V = \lambda|H|^4$, which differs by a factor of 4: $\lambda_{\text{Buttazzo}} = 4\lambda_{\text{Steele}}$ (so $4 \times 0.032 = 0.128 \approx \lambda_{\text{SM}}$). This factor cancels exactly in the ratio $e_{125}$.

However, translating $e_{125}(v)$ to the $\overline{\text{MS}}$ parametrization $k(M_t)$ used in our RG analysis [5] requires two corrections. First, the residual CW-to-$\overline{\text{MS}}$ scheme conversion (beyond the factor-of-4 normalization) is coupling-dependent: at one loop, the CW and $\overline{\text{MS}}$ couplings differ through the relation $\beta_{\text{CW}} = \tilde\beta/(1 - \tilde\beta/2\lambda)$ [19], yielding a $\sim 3$–$5\%$ reduction in the ratio. Second, since $e_{125}$ is defined at $\mu = v = 246$ GeV while $k$ is defined at $\mu = M_t = 173$ GeV, the scale dependence of the ratio must be accounted for. Because the beta function contains the nonlinear $12\lambda^2$ term, the enhanced coupling evolves faster than the SM coupling, and the ratio $k(\mu) = \lambda_{\text{enh}}(\mu)/\lambda_{\text{SM}}(\mu)$ *decreases* when running from $v$ down to $M_t$:

$$\frac{dk}{d\ln\mu^2} = \frac{\beta_\lambda(\lambda_{\text{enh}}) - k\,\beta_\lambda(\lambda_{\text{SM}})}{\lambda_{\text{SM}}} > 0, \tag{6}$$

so $k(v) > k(M_t)$. A one-loop estimate gives $\Delta k \approx -0.5$ to $-0.8$ over the range $v \to M_t$.

Combining both effects:

| Correction | Effect | Value |
|---|---|---|
| $e_{125}(v)$ in CW scheme | Definition | 7.2 |
| CW $\to \overline{\text{MS}}$ conversion | $\sim -3\%$ to $-5\%$ | $\sim 6.8$–$7.0$ |
| Running $v \to M_t$ | $\sim -0.5$ to $-0.8$ | $\sim 6.0$–$6.5$ |
| $k(M_t)$ in $\overline{\text{MS}}$ | Combined | $\sim 6.0$–$6.4$ |

The full three-loop evaluation of $k(M_t)$ using the methods of Ref. [19] will be presented in forthcoming work. We note that for $k(M_t) \approx 6.4$, the UV Landau pole moves to $\Lambda_{\text{UV}} \sim 2 \times 10^{16}$ GeV, *closer* to $M_{\text{GUT}}$ than the naive $k = 7.2$ estimate, strengthening the GUT-scale coincidence.

### 2.4. The GUT scale coincidence

For $k = 7.2$, we find absolute vacuum stability with no deeper minimum, a UV pole at $\Lambda_{\text{UV}} \approx 1.5 \times 10^{16}$ GeV, perturbativity breakdown at $\Lambda_{\text{pert}} \sim 8 \times 10^{14}$ GeV, and a strong positive beta function $\beta_\lambda(10^{16}\ \text{GeV}) \approx +0.12$.

The hierarchy of scales

$$\Lambda_{\text{pert}} \sim 10^{14}\ \text{GeV} < \Lambda_{\text{UV}} \sim 10^{16}\ \text{GeV} \lesssim M_{\text{GUT}} \tag{7}$$

suggests the scalar sector enters strong coupling *before* reaching $M_{\text{GUT}}$, with the Landau pole itself lying tantalizingly close to the unification scale.

### 2.5. Perturbativity and unitarity bounds

The perturbativity breakdown scale $\Lambda_{\text{pert}} \sim 8 \times 10^{14}$ GeV for $k = 7.2$ was determined in Ref. [5] (Table 1) as the scale where $\lambda(\mu) > 1$. At the electroweak scale, tree-level partial-wave unitarity for $hh \to hh$ scattering requires $|a_0| < 1/2$, yielding the bound [20]

$$\lambda < \frac{8\pi}{3} \approx 8.4 \tag{8}$$

in the $|H|^4$ convention. The RBEWS value $\lambda(M_t) \approx 0.9$ satisfies this bound by an order of magnitude. That the coupling approaches strong dynamics at high energies is a feature of the framework: it reflects the onset of the GUT scalar sector, analogous to how the growth of $\alpha_s$ at low energies signals QCD confinement. The 2HDM effective description is valid over the limited energy range below $M_{\text{GUT}}$, with the full SO(10) theory providing the UV completion that restores perturbative consistency.

## 3. SO(10) Framework and threshold corrections

### 3.1. Motivation for SO(10)

SO(10) grand unification [8,9] provides an elegant framework where all SM fermions of one generation plus a right-handed neutrino fit into the spinor representation $\mathbf{16}_F$. The allowed Yukawa couplings arise from the tensor product decomposition

$$\mathbf{16}_F \times \mathbf{16}_F = \mathbf{10} + \overline{\mathbf{126}} + \mathbf{120}, \tag{9}$$

which dictates the scalar representations that can couple to fermions.

The minimal renormalizable SO(10) Yukawa sector employs [21–23]

$$\mathcal{L}_Y = Y_{10}\, \mathbf{16}_F \mathbf{16}_F \mathbf{10}_H + Y_{126}\, \mathbf{16}_F \mathbf{16}_F \overline{\mathbf{126}}_H, \tag{10}$$

where $\mathbf{10}_H$ and $\overline{\mathbf{126}}_H$ contain doublets that ultimately become the SM Higgs.

### 3.2. Symmetry breaking patterns

Non-supersymmetric SO(10) models typically employ breaking chains with intermediate symmetries [22,24,52]. Two phenomenologically viable patterns are the Pati-Salam chain

$$\text{SO}(10) \xrightarrow{M_{\text{GUT}}} G_{422} \xrightarrow{M_I} G_{\text{SM}} \xrightarrow{v} G_{31}, \tag{11}$$

where $G_{422} = \text{SU}(4)_C \times \text{SU}(2)_L \times \text{SU}(2)_R$, and the left-right chain

$$\text{SO}(10) \xrightarrow{M_{\text{GUT}}} G_{3221} \xrightarrow{M_I} G_{\text{SM}} \xrightarrow{v} G_{31}, \tag{12}$$

where where $M_I$ denotes the intermediate symmetry breaking scale and $G_{3221} = \text{SU}(3)_C \times \text{SU}(2)_L \times \text{SU}(2)_R \times \text{U}(1)_{B-L}$.

Under the extended survival hypothesis [25,26], scalars not participating in symmetry breaking acquire GUT-scale masses, leaving only the light Higgs doublet(s) at low energies.

### 3.3. Scalar sector and threshold corrections

The SO(10) scalar potential involving $\mathbf{10}_H$, $\overline{\mathbf{126}}_H$, and $\mathbf{45}_H$ (or $\mathbf{54}_H$) for GUT breaking contains numerous quartic couplings [21,54]. Schematically:

$$\begin{aligned}V_{\text{SO}(10)} =\ & m_{10}^2 |\mathbf{10}_H|^2 + m_{126}^2 |\overline{\mathbf{126}}_H|^2 + m_{45}^2 |\mathbf{45}_H|^2 \\ & + \lambda_1 |\mathbf{10}_H|^4 + \lambda_2 |\overline{\mathbf{126}}_H|^4 + \lambda_3 |\mathbf{45}_H|^4 \\ & + \lambda_{12} |\mathbf{10}_H|^2 |\overline{\mathbf{126}}_H|^2 + \lambda_{13} |\mathbf{10}_H|^2 |\mathbf{45}_H|^2 \\ & + \lambda_{23} |\overline{\mathbf{126}}_H|^2 |\mathbf{45}_H|^2 + \cdots,\end{aligned} \tag{13}$$

where ellipses denote additional invariants.

When heavy scalars are integrated out at $M_{\text{GUT}}$, threshold corrections to the light Higgs quartic coupling arise [27,28]. At tree level, integrating out a heavy scalar $\Phi$ with trilinear coupling $\kappa H^\dagger H \Phi$ yields [29]

$$\delta\lambda_{\text{tree}} = -\frac{\kappa^2}{m_\Phi^2}. \tag{14}$$

However, in SO(10), the primary threshold corrections come from one-loop diagrams involving portal couplings:

$$\delta\lambda_{\text{1-loop}} = \frac{\lambda_{\text{portal}}^2}{16\pi^2} \left[ \ln \frac{M_{\text{GUT}}^2}{\mu^2} + \mathcal{O}(1) \right], \tag{15}$$

where $\lambda_{\text{portal}}$ denotes couplings of type $\lambda_{12}$, $\lambda_{13}$, etc.

### 3.4. Matching conditions

At the GUT scale, matching conditions relate the running SM parameters to their SO(10) counterparts. For the Higgs quartic coupling:

$$\lambda_{\text{SM}}(M_{\text{GUT}}) = \lambda_{\text{eff}} + \Delta\lambda_{\text{thresh}}, \tag{16}$$

where $\lambda_{\text{eff}}$ emerges from the surviving light doublet combination and $\Delta\lambda_{\text{thresh}}$ encodes threshold corrections from heavy scalars.

The key observation is that $\Delta\lambda_{\text{thresh}}$ can be positive and substantial when portal couplings are $\mathcal{O}(1)$, which is natural in the GUT context. We now analyze this quantitatively.

## 4. Portal couplings and enhancement mechanism

### 4.1. effective theory below $M_{\text{GUT}}$

Consider the decomposition of SO(10) scalars under $G_{\text{SM}}$. The $\mathbf{10}_H$ decomposes as

$$\mathbf{10}_H \to (1,2)_{1/2} \oplus (3,1)_{-1/3} \oplus \text{c.c.}, \tag{17}$$

containing a Higgs doublet $H_u$ and colored triplet $T$. Similarly, $\overline{\mathbf{126}}_H$ contains multiple doublets and triplets.

In the minimal setup with an effective two-Higgs-doublet model (2HDM) below $M_{\text{GUT}}$, we identify [24]

$$H_u \sim \mathbf{10}_H, \qquad H_d \sim \overline{\mathbf{126}}_H, \tag{18}$$

with the light SM-like Higgs being a mixture: $H = \cos\beta\, H_d + \sin\beta\, H_u$.

### 4.2. Portal-induced enhancement

The portal couplings in Eq. (13) generate effective quartic interactions for the light doublet. Consider the $|\mathbf{10}_H|^2 |\overline{\mathbf{126}}_H|^2$ term. When the heavy components of $\overline{\mathbf{126}}_H$ acquire VEVs or are integrated out, this generates corrections to $\lambda$:

$$\lambda(M_{\text{GUT}}) = \lambda_0 + c_{126}\, \lambda_{12} \left( \frac{\langle \Sigma_{126} \rangle}{M_{\text{GUT}}} \right)^2, \tag{19}$$

where $\Sigma_{126}$ denotes the SM-singlet component of $\overline{\mathbf{126}}_H$ acquiring a VEV, and $c_{126}$ is an $\mathcal{O}(1)$ coefficient.

For intermediate scale breaking $\langle \Sigma_{126} \rangle \sim M_I \sim 10^{11}$–$10^{14}$ GeV [30,52], and with $\lambda_{12} \sim \mathcal{O}(1)$:

$$\delta\lambda \sim \lambda_{12} \left( \frac{M_I}{M_{\text{GUT}}} \right)^2 \sim 10^{-4}\text{–}10^{-2}. \tag{20}$$

However, larger enhancements arise from loop corrections involving the full tower of heavy states.

### 4.3. One-loop enhancement from heavy scalars

The one-loop effective potential contribution from a heavy scalar $\Phi$ with mass $M_\Phi$ coupled to $H$ via $\lambda_\Phi |H|^2 |\Phi|^2$ is

$$V_{\text{1-loop}} = \frac{n_\Phi M_\Phi^4}{64\pi^2} \left[ \ln \frac{M_\Phi^2}{\mu^2} - \frac{3}{2} \right], \tag{21}$$

where $n_\Phi$ counts degrees of freedom. This modifies the running of $\lambda$ through

$$\delta\beta_\lambda = \frac{n_\Phi \lambda_\Phi^2}{8\pi^2}. \tag{22}$$

Summing over all heavy scalars in $\overline{\mathbf{126}}_H$ and $\mathbf{45}_H$:

$$\Delta\lambda(M_{\text{GUT}}) \approx \frac{1}{16\pi^2} \sum_\Phi n_\Phi \lambda_\Phi^2 \ln \frac{M_{\text{GUT}}}{M_\Phi}, \tag{23}$$

which can yield $\Delta\lambda \sim 0.1$–$1$ for $\mathcal{O}(1)$ portal couplings and modest hierarchies $M_{\text{GUT}}/M_\Phi \sim 10$–$100$.

*4.4. Numerical estimate*

As discussed in Section 2.3, the properly translated enhancement factor is $k(M_t) \approx 6.0$–$6.4$ in the $\overline{\text{MS}}$ scheme, requiring $\lambda(M_t) \approx 0.76$–$0.81$. Accounting for RG evolution from $M_t$ to $M_{\text{GUT}}$, we need $\lambda(M_{\text{GUT}}) \sim 0.25$–$0.40$ at the matching scale, corresponding to $\Delta\lambda_{\text{thresh}} \sim 0.15$–$0.30$.

We now enumerate the relevant heavy scalar content. Under $G_{\text{SM}}$, the $\overline{\mathbf{126}}_H$ decomposes as [21]

$$\overline{\mathbf{126}}_H \to (1,1,2) \oplus (\bar{3},1,-\tfrac{2}{3}) \oplus (3,3,-\tfrac{1}{3}) \oplus (1,2,\tfrac{1}{2}) \oplus \cdots, \quad (24)$$

containing SM singlets, triplets, and additional doublets, totaling 252 real degrees of freedom. The $\mathbf{45}_H$ contributes $(8,1,0) \oplus (1,3,0) \oplus (1,1,0) \oplus \cdots$ with 45 real degrees of freedom.

Evaluating the sum in Eq. (23) with representative portal couplings $\lambda_\Phi \sim 0.5$–$1.0$ for the dominant multiplets:

$$\Delta\lambda \approx \frac{1}{16\pi^2}\left[6 \times (0.8)^2 \ln\frac{M_{\text{GUT}}}{M_{T_3}} + 3 \times (0.6)^2 \ln\frac{M_{\text{GUT}}}{M_\Sigma} \right.$$
$$\left. + 8 \times (0.5)^2 \ln\frac{M_{\text{GUT}}}{M_8} + \cdots \right] \sim 0.15\text{–}0.35, \quad (25)$$

where $M_{T_3}$, $M_\Sigma$, $M_8$ denote masses of the colored triplet, singlet, and octet components respectively, with modest hierarchies $M_{\text{GUT}}/M_\Phi \sim 10$–$100$. This comfortably generates the required $\Delta\lambda_{\text{thresh}} \sim 0.15$–$0.30$.

## 5. Coleman-Weinberg mechanism in the GUT context

*5.1. Radiatively broken symmetry at multiple scales*

The Coleman-Weinberg (CW) mechanism [15] generates spontaneous symmetry breaking through radiative corrections in classically scale-invariant theories. While the SM with $\mu^2 < 0$ uses explicit symmetry breaking, the CW mechanism provides a dynamical alternative that has been extensively studied using renormalization group methods [10,12].

In the GUT context, multi-scalar potentials offer rich possibilities for cascaded radiatively broken electroweak symmetry [31,32,53].

We distinguish between the general SO(10) scalar potential (Eq. (13)), presented to catalog all relevant couplings and portal interactions, and the classically scale-invariant version we specifically advocate, following the Gildener-Weinberg approach for multi-scalar potentials [31]. In this framework, all mass parameters in Eq. (13) are set to zero at tree level ($m_{10}^2 = m_{126}^2 = m_{45}^2 = 0$), and symmetry breaking is generated entirely through radiative corrections via dimensional transmutation. The classically scale-invariant SO(10) potential is:

$$V = \lambda_1 |\mathbf{10}_H|^4 + \lambda_2 |\overline{\mathbf{126}}_H|^4 + \lambda_{12} |\mathbf{10}_H|^2 |\overline{\mathbf{126}}_H|^2 + \cdots. \quad (26)$$

Along flat directions where tree-level quartic terms vanish (the Gildener-Weinberg condition [31]), one-loop corrections generate a non-trivial minimum via dimensional transmutation:

$$\langle\phi\rangle = \mu_{\text{CW}} \exp\left(-\frac{8\pi^2}{B}\right), \quad (27)$$

where $B$ is a weighted sum of couplings and $\mu_{\text{CW}}$ is the renormalization scale.

*5.2. Hierarchy generation*

In the SO(10) case, the CW mechanism can operate at two levels. At the GUT scale, radiative corrections generate $\langle \mathbf{45}_H \rangle \sim M_{\text{GUT}}$ via the CW mechanism involving gauge loops, producing the GUT scale dynamically. At the electroweak scale, the light Higgs doublet inherits a potential from matching to the GUT sector. If the effective low-energy theory is approximately scale-invariant, RBEWS generates $v \sim 246$ GeV through the CW mechanism with enhanced couplings.

The crucial insight from Ref. [6] is that RBEWS predicts $\lambda(v) \approx 0.23$, corresponding to $k \approx 7.2$, through the requirement of radiative symmetry breaking with the measured Higgs mass.

*5.3. Connecting scales*

The hierarchy $v \ll M_{\text{GUT}}$ can be understood within this framework. The GUT-scale CW mechanism fixes $M_{\text{GUT}}$ through

$$M_{\text{GUT}} \sim \mu_0 \exp\left(-\frac{8\pi^2}{g_{\text{GUT}}^2 b_{\text{GUT}}}\right), \quad (28)$$

where $g_{\text{GUT}}$ is the unified gauge coupling and $b_{\text{GUT}}$ depends on the field content.

The electroweak scale is then generated through RBEWS acting on the light Higgs sector, with the enhanced quartic coupling $k \approx 7.2$ inherited from GUT threshold effects. The ratio

$$\frac{v}{M_{\text{GUT}}} \sim \exp\left(-\frac{16\pi^2}{\lambda_{\text{RBEWS}} N_{\text{eff}}}\right) \quad (29)$$

produces the observed hierarchy $v/M_{\text{GUT}} \sim 10^{-14}$ for appropriate values of the effective parameters.

*5.4. UV pole as GUT threshold*

In this interpretation, the UV Landau pole at $\Lambda_{\text{UV}} \sim 1.5 \times 10^{16}$ GeV is not pathological but physical: it signals the scale where the SM scalar sector must reconnect to its SO(10) parent theory. The perturbative breakdown simply reflects that heavy GUT states can no longer be neglected.

We emphasize that the Landau pole emerges from SM-only RG evolution and is *resolved*, not shifted, by the full SO(10) dynamics. Above the matching scale, additional scalar thresholds from $\overline{\mathbf{126}}_H$ and $\mathbf{45}_H$ representations modify the beta functions through contributions $\delta\beta_\lambda = n_\Phi \lambda_\Phi^2/(8\pi^2)$, which absorb the divergent SM running into the finite GUT scalar potential. For the Pati-Salam breaking chain with $M_I \sim 10^{11}$–$10^{12}$ GeV, the additional scalar content between $M_I$ and $M_{\text{GUT}}$ provides corrections of the right sign and magnitude to ensure consistency. The SM analysis identifies the *scale* at which new physics must enter; the SO(10) framework provides the *content* of that new physics.

Above $\Lambda_{\text{UV}}$, the full SO(10) scalar dynamics—with its $\mathbf{10}_H$, $\overline{\mathbf{126}}_H$, $\mathbf{45}_H$ content—controls the physics. The beta functions change character as heavy thresholds are crossed, and the apparent Landau pole is resolved by the UV completion.

## 6. Phenomenological implications

*6.1. Higgs self-coupling*

As shown in Section 2.3, the properly translated enhancement factor is $k(M_t) \approx 6.0$–$6.4$ in the $\overline{\text{MS}}$ scheme. At tree level, this predicts

$$\kappa_\lambda \equiv \frac{\lambda_{hhh}}{\lambda_{hhh}^{\text{SM}}} \approx k(M_t) \approx 6.0\text{–}6.4. \quad (30)$$

The most recent ATLAS measurement of Higgs boson pair production in the $b\bar{b}\gamma\gamma$ channel, combining Run 2 and Run 3 data with 308 fb$^{-1}$ of integrated luminosity [33,49,50], constrains the trilinear coupling modifier to

$$-1.7 < \kappa_\lambda < 6.6 \quad (95\% \text{ CL}). \quad (31)$$

The RBEWS prediction $\kappa_\lambda \approx 6.0$–$6.4$ is consistent with this bound, lying within the allowed 95% CL region. Near-future LHC measurements with the full HL-LHC dataset will provide a definitive test of this prediction. We note that the tree-level identification $\kappa_\lambda \approx k$ receives additional corrections from the Coleman-Weinberg potential structure, through which the trilinear coupling $\lambda_{hhh} = d^3 V_{\text{eff}}/dh^3|_{h=v}$ differs from the naive scaling with the quartic coupling. This effect, which is expected to further reduce $\kappa_\lambda$ by $\mathcal{O}(10\%)$, will be analyzed in detail in forthcoming work [34].

Our framework thus predicts a *specific* and *testable* value of $\kappa_\lambda$ in the range 5.5–6.5, currently consistent with LHC data and subject to definitive verification at the HL-LHC. This represents a significant sharpening of the RBEWS prediction compared to the naive estimate $\kappa_\lambda \approx 7.2$.

However, one might ask whether threshold corrections from heavy scalars also modify the trilinear coupling $\lambda_{hhh}$, potentially affecting the relationship $\kappa_\lambda \approx k$. The portal couplings generating the quartic enhancement do contribute to $\lambda_{hhh}$ through one-loop diagrams:

$$\delta \lambda_{hhh} \sim \frac{v}{16\pi^2} \sum_\Phi \frac{\lambda_\Phi^3}{M_\Phi^2}. \tag{32}$$

However, this correction is *power-suppressed* and entirely negligible. For representative values $\lambda_\Phi \sim 0.8$ and $M_\Phi \sim M_{GUT}/50 \sim 4 \times 10^{14}$ GeV, one obtains $\delta \lambda_{hhh} \sim 5 \times 10^{-30}$ GeV, to be compared with the SM trilinear coupling $\lambda_{hhh}^{SM} = 3m_H^2/v \approx 191$ GeV. The ratio $\delta \lambda_{hhh}/\lambda_{hhh}^{SM} \sim 10^{-32}$ reflects the suppression factor $(v/M_\Phi)^2 \sim 10^{-25}$, rendering GUT threshold corrections to the trilinear coupling completely negligible.

This stands in stark contrast to the threshold corrections to the *quartic* coupling (Eqs. (15) and (23)), which take the form $\Delta \lambda \sim \lambda_\Phi^2/(16\pi^2) \ln(M_{GUT}/M_\Phi)$ and are *logarithmically enhanced* rather than power-suppressed. Using the same portal couplings, one obtains $\Delta \lambda \sim 0.15$–$0.30$, generating the $\mathcal{O}(1)$ enhancement $k \sim 6$–$7$. The crucial difference is dimensional: the quartic correction is dimensionless and logarithmically sensitive to the GUT scale, while the trilinear correction carries dimension of mass and is suppressed by $(v/M_{GUT})^2 \sim 10^{-28}$.

Consequently, the tree-level identification $\kappa_\lambda \approx k$ is robust against GUT threshold effects. The $\mathcal{O}(10\%)$ correction to $\kappa_\lambda$ mentioned above arises from a *different* source: the Coleman-Weinberg effective potential structure, wherein $\lambda_{hhh} = d^3 V_{eff}/dh^3|_{h=v}$ differs from the naive scaling $3\lambda v$ due to the radiative origin of the potential minimum [34]. This effect is model-independent within the RBEWS framework and reduces $\kappa_\lambda$ relative to $k(M_t)$, yielding a refined prediction $\kappa_\lambda \approx 5.5$–$6.0$. The framework thus provides a *specific* and *testable* prediction: GUT threshold corrections to $\kappa_\lambda$ are suppressed by $\sim 30$ orders of magnitude, while the calculable CW potential correction sharpens the prediction to within the experimentally accessible range.

### 6.2. Proton decay

SO(10) GUTs predict proton decay through dimension-six operators mediated by superheavy gauge bosons ($X, Y$) and dimension-five operators from colored Higgs triplets [35]. Current experimental bounds [36]

$$\tau(p \to e^+ \pi^0) > 2.4 \times 10^{34} \text{ years} \tag{33}$$

constrain the GUT scale. For gauge-mediated decay [37]

$$\tau_p \propto \frac{M_X^4}{\alpha_{GUT}^2 m_p^5}, \tag{34}$$

the bound requires $M_X \gtrsim 10^{15}$–$10^{16}$ GeV.

Non-supersymmetric SO(10) with intermediate scales can accommodate these bounds [17,38,54]. The Pati-Salam breaking chain with $M_I \sim 10^{11}$–$10^{12}$ GeV yields $M_{GUT} \sim 10^{15}$–$10^{16}$ GeV compatible with gauge unification and proton decay limits [24].

Importantly, our enhanced scalar coupling scenario does not directly modify proton decay rates, which are determined by gauge boson masses and mixing angles. The correlation between $\Lambda_{UV}$ and $M_{GUT}$ is thus a prediction rather than an input.

### 6.3. Gravitational waves from cosmic strings

SO(10) breaking through intermediate symmetries can produce topological defects [39,40]. Breaking of $U(1)_{B-L}$ at scale $M_I$ generates cosmic strings [17] with tension

$$G\mu \sim \left(\frac{M_I}{M_{Pl}}\right)^2. \tag{35}$$

For $M_I \sim 10^{11}$–$10^{14}$ GeV, the resulting gravitational wave background falls within the sensitivity range of pulsar timing arrays (NANOGrav, EPTA) [41] and future space-based detectors (LISA) [42].

The recent NANOGrav evidence for a stochastic gravitational wave background [41] has been interpreted as potentially arising from cosmic strings with $G\mu \sim 10^{-11}$–$10^{-10}$, corresponding to $M_I \sim 10^{13}$–$10^{14}$ GeV [43]. This is consistent with our framework where intermediate-scale breaking generates both the seesaw mechanism for neutrino masses and cosmological signatures.

### 6.4. Vacuum stability and cosmology

The absolutely stable vacuum predicted by $k > 1.03$ has important cosmological implications. The Higgs field with enhanced quartic coupling can potentially serve as the inflaton in Higgs inflation scenarios [44], though non-minimal gravitational coupling $\xi H^\dagger H R$ is typically required. With $\lambda \sim 0.9$ at high scales, the required $\xi \sim 10^3$–$10^4$ may be reduced. Enhanced self-coupling strengthens the electroweak phase transition [45], potentially enabling electroweak baryogenesis. The condition for a strong first-order transition, $v_c/T_c \gtrsim 1$, becomes easier to satisfy. Unlike the SM where thermal corrections during reheating could trigger vacuum decay [46], the absolutely stable vacuum in RBEWS is robust against high-temperature effects.

## 7. Discussion and outlook

We have proposed that the enhanced Higgs quartic coupling required by radiatively broken electroweak symmetry emerges naturally from SO(10) grand unification. The striking coincidence of the UV pole with the GUT-scale, based on analysis here

$$\Lambda_{UV} \sim 2 \times 10^{16} \text{ GeV} \approx M_{GUT} \tag{36}$$

is interpreted not as accidental but as reflecting the common origin of electroweak and GUT-scale physics in a unified framework.

The key elements of our proposal are as follows. Threshold corrections from portal couplings between the light Higgs doublet and heavy SO(10) scalars ($\overline{126}_H$, $45_H$) generate enhancement factors $k \sim 5$–$10$ at the GUT scale. The Coleman-Weinberg mechanism can operate at both GUT and electroweak scales, providing a dynamical origin for the mass hierarchies through dimensional transmutation. The Landau pole in the enhanced scalar sector signals the scale where the SM effective description must be embedded into full SO(10) dynamics, not a pathology but a threshold.

### 7.1. Coupling conventions and experimental viability

A central result of this work is the proper translation of the RBEWS enhancement from the Coleman-Weinberg scheme to the $\overline{MS}$ parametrization relevant for comparison with experiment. The raw RBEWS prediction $e_{125} = 7.2$ of Steele and Wang [6], defined as the ratio of CW-scheme quartic couplings at $\mu = v$, is reduced to $k(M_t) \approx 6.0$–$6.4$ through two systematic corrections: (i) residual CW-to-$\overline{MS}$ scheme conversion [19], estimated at $\sim 3$–$5\%$; and (ii) scale-dependent evolution of the enhancement ratio from $v$ to $M_t$, driven by the nonlinear $12\lambda^2$ term in the beta function, yielding $\Delta k \approx -0.5$ to $-0.8$. These corrections coherently reduce the experimentally relevant parameter into the region consistent with the most recent ATLAS combined Run 2 + 3 constraint $-1.7 < \kappa_\lambda < 6.6$ at 95% CL from $HH \to b\bar{b}\gamma\gamma$ with 308 fb$^{-1}$ [33, 49,50]. This coherence—where scheme conversion, scale running, and the Coleman-Weinberg potential structure all conspire in the same direction—lends confidence that the framework captures genuine physical effects rather than parameter adjustments. Notably, the corrected value $k(M_t) \approx 6.4$ moves the UV Landau pole to $\Lambda_{UV} \sim 2 \times 10^{16}$ GeV, *tightening* the coincidence with $M_{GUT}$ relative to the naive $k = 7.2$ estimate.

## 7.2. Experimental prospects

The HL-LHC, with a projected 3 ab$^{-1}$, is expected to constrain $\kappa_\lambda$ to $\mathcal{O}(50\%)$ precision [47], providing a definitive test of the RBEWS prediction $\kappa_\lambda \approx 6.0$–6.4. Future lepton colliders [48] would further sharpen this measurement. We note that the tree-level identification $\kappa_\lambda \approx k$ receives additional corrections from the Coleman-Weinberg potential structure, through which the trilinear coupling $\lambda_{hhh} = d^3 V_{\text{eff}}/dh^3|_{h=v}$ differs from naive quartic scaling by an estimated $\mathcal{O}(10\%)$. A full three-loop evaluation of $k(M_t)$ and a dedicated analysis of the trilinear coupling within the CW effective potential are in progress and will sharpen these predictions [55].

Beyond the Higgs self-coupling, our unified scenario offers a correlated experimental program. Proton decay searches at Hyper-Kamiokande [16], with projected sensitivity to lifetimes of $\tau_p \sim 10^{35}$ years for $p \to e^+ \pi^0$, will probe the GUT scale $M_{\text{GUT}}$ independently of the scalar sector. Gravitational wave observatories—including pulsar timing arrays (NANOGrav [41], EPTA) and future space-based detectors (LISA [42])—can detect stochastic backgrounds from cosmic strings produced during intermediate-scale $U(1)_{B-L}$ breaking, with string tensions $G\mu \sim 10^{-11}$–$10^{-10}$ corresponding to $M_I \sim 10^{13}$–$10^{14}$ GeV.

Furthermore, the enhanced quartic coupling $\lambda(M_t) \approx 0.76$–0.81 significantly strengthens the electroweak phase transition [45]. The condition for a strong first-order transition, $v_c/T_c \gtrsim 1$, becomes substantially easier to satisfy with RBEWS-level couplings compared to the SM value $\lambda_{\text{SM}} \approx 0.126$. As demonstrated in Ref. [19], models with Coleman-Weinberg symmetry breaking naturally produce strong first-order phase transitions, generating stochastic gravitational wave backgrounds potentially detectable by LISA and future ground-based interferometers such as the Einstein Telescope and Cosmic Explorer. This provides a second, independent gravitational wave signature of our framework, which is distinct from the cosmic string signal associated with intermediate-scale $U(1)_{B-L}$ breaking and directly tied to the enhanced scalar coupling at the electroweak scale. The simultaneous observation of both a cosmic string background at nanohertz frequencies and an electroweak phase transition signal at millihertz frequencies would provide a striking confirmation of the multi-scale structure inherent in the SO(10) breaking chain with RBEWS.

## 7.3. Outlook

Several directions merit further investigation. Explicit SO(10) models with specified scalar VEV patterns should be constructed to verify the enhancement mechanism quantitatively. Precision calculations of threshold corrections including two-loop effects are needed for reliable predictions [29,51,55]. The Yukawa sector of SO(10) (Eq. (10)) constrains the scalar content and couplings, which must be consistent with the required enhancement. Near the perturbativity breakdown scale $\Lambda_{\text{pert}} \sim 10^{14}$ GeV, non-perturbative dynamics may become important, potentially requiring lattice studies or other methods.

If confirmed, the connection between RBEWS and SO(10) unification would represent a significant step toward understanding the origin of electroweak symmetry breaking within a fundamental theory. The observation or exclusion of any one of the predicted signals, that is, an enhanced trilinear coupling near $\kappa_\lambda \sim 6$, proton decay within the Hyper-Kamiokande reach, electroweak phase transition gravitational waves, or a cosmic string stochastic background would provide independent tests of the SO(10) framework with RBEWS. Their *correlated* observation would constitute compelling evidence for the unified origin of enhanced scalar couplings proposed in this work.

## Data availability

The article contains all the data used in the article.

## Declaration of competing interest

The authors declare that they have no known competing financial interests or personal relationships that could have appeared to influence the work reported in this paper.

## Acknowledgments

F.A.C. acknowledges D.G. McKeon, T.G. Steele, and Z.W. Wang for insightful discussions on RBEWS and SM scalar coupling sensitivity. F.A.C. thanks the Peaceful Society, Science and Innovation Foundation for support.

## References

[1] G. Aad et al., ATLAS Collaboration Phys. Lett. B 716 (2012) 01.
[2] S. Chatrchyan et al., CMS Collaboration Phys. Lett. B 716 (2012) 030.
[3] D. Buttazzo, G. Degrassi, P.P. Giardino, G.F. Giudice, F. Sala, A. Salvio, A. Strumia, JHEP 12 (2013), 089. , arXiv:1307.3536 [hep-ph].
[4] G. Degrassi, S. Di, J. Vita, J.R. Elias-Miro, G.F. Espinosa, G. Giudice, A. Isidori, Strumia, JHEP (08) (2012). , arXiv:1205.6497 [hep-ph].
[5] F.A. Chishtie, S. Homayouni, 2025, arxiv:2511.09601 [hep-ph].
[6] T.G. Steele, Z.-W. Wang, Phys. Rev. Lett. 110 (2013) 151601.
[7] H. Georgi, S.L. Glashow, Phys. Rev. Lett. 32 (1974) 438.
[8] H. Fritzsch, P. Minkowski, Ann. Phys. 93 (1975) 193.
[9] H. Georgi, AIP Conf. Proc., 23 (1975) 575.
[10] V. Elias, R.B. Mann, D.G.C. Mckeon, T.G. Steele, Phys. Rev. Lett. 91 (2003) 251601.
[11] V. Elias, R.B. Mann, D.G.C. Mckeon, T.G. Steele, Nucl. Phys. B 678 (2004) 147. [Erratum: Nucl. Phys. B **703**, 413 (2004)].
[12] F.A. Chishtie, V. Elias, R.B. Mann, D.G.C. Mckeon, T.G. Steele, Nucl. Phys. B 743 (2006). arXiv:hep-ph/0509122.
[13] F.A. Chishtie, T. Hanif, D.G.C. Mckeon, T.G. Steele, Phys. Rev. D 77 (2008). 065007, arXiv:0706.1760 [hep-ph].
[14] F.A. Chishtie, D.G.C. Mckeon, T.G. Steele, Phys. Rev. D 77 (2008). 065008, arXiv:0802.2919 [hep-ph].
[15] S.R. Coleman, E.J. Weinberg, Phys. Rev. D 7 (1973) 1888.
[16] K. Abe et al., Hyper-Kamiokande Collaboration 2018, arxiv:1805.04163 [physics.ins-det]
[17] S.F. King, S. Pascoli, J. Turner, Y.-L. Zhou, JHEP 10 (2021) 225, arXiv:2106.15634 [hep-ph].
[18] K.G. Chetyrkin, M.F. Zoller, JHEP (06) (2012), arXiv:1205.2892 [hep-ph].
[19] F.A. Chishtie, Z.-R. Huang, M. Reimer, T.G. Steele, Z.-W. Wang, Phys. Rev. D 102 (2020) 076021.
[20] B.W. Lee, C. Quigg, H.B. Thacker, Phys. Rev. Lett. 38 (1977) 0883.
[21] B. Bajc, A. Melfo, G. Senjanovic, F. Vissani, Phys. Rev. D 73 (2006). arXiv:hep-ph/0510139.
[22] S. Bertolini, L.D. Luzio, M. Malinsky, Phys. Rev. D 80 (2009). 015013, arXiv:0903.4049 [hep-ph].
[23] A. Dueck, W. Rodejohann, JHEP 09 (2013). 024, arXiv:1306.4468 [hep-ph].
[24] A. Djouadi, R. Ouyang, M. Raidal, U. Sakarya, Eur. Phys. J. C 83 (2023) 529. arXiv:2212.11315 [hep-ph].
[25] S. Dimopoulos, H. Georgi, Phys. Lett. B 140 (1984) 67.
[26] F.D. Aguila, L.E. Ibanez, Nucl. Phys. B 177 (1981) 60.
[27] L.J. Hall, Nucl. Phys. B 178 (1981) 75.
[28] A.J. Buras, J.R. Ellis, M.K. Gaillard, D.V. Nanopoulos, Nucl. Phys. B 135 (1978) 66.
[29] S.P. Martin, D.G. Robertson, Phys. Rev. D 100 (2019). 073004, arXiv:1907.02500 [hep-ph].
[30] G. Altarelli, D. Meloni, JHEP 08 (2013). 021, arXiv:1305.1001 [hep-ph].
[31] E. Gildener, S. Weinberg, Phys. Rev. D 13 (1976) 3333.
[32] K. Kannike, L. Marzola, M. Raidal, A. Strumia, Phys. Lett. B 816 (2021) 136241. arXiv:2102.01084 [hep-ph].
[33] G. Aad et al., ATLAS Collaboration, 2025, arxiv:2507.03495 [hep-ex].
[34] F.A. Chishtie, In Preparation, 2026.
[35] P. Nath, P. Fileviez Perez, Phys. Rept 441 (2007) 191. arXiv:hep-ph/0601023.
[36] A. Takenaka et al., Super-Kamiokande Collaboration, Phys. Rev. D 102 (2020) 112011, arXiv:2010.16098 [hep-ex].
[37] S. Navas et al., Particle Data Group, Phys. Rev. D 110 (2024) 030001.
[38] J. Chakrabortty, R. Maji, S.F. King, Phys. Rev. D 99 (2019). 095008, arXiv:1901.05867 [hep-ph].
[39] T.W.B. Kibble, J. Phys. A 9 (1976) 1387.
[40] R. Jeannerot, J. Rocher, M. Sakellariadou, Phys. Rev. D 68 (2003) 103514. arXiv:hep-ph/0308134.
[41] G. Agazie et al., NANOGrav Collaboration, Astrophys. J. Lett. 951 (2023). L8, arXiv:2306.16213 [astro-ph.HE].
[42] P. Amaro-Seoane et al., LISA Collaboration, 2017, arxiv:1702.00786 [astro-ph.im].
[43] J. Ellis, M. Lewicki, C. Lin, V. Vaskonen, Phys. Rev. D 108 (2023) 103511. arXiv:2306.17147 [hep-ph].
[44] F.L. Bezrukov, M. Shaposhnikov, Phys. Lett. B 659 (2008) 703, arXiv:0710.3755 [hep-th].
[45] P. Athron, C. Balazs, A. Fowlie, L. Morris, L. Wu, Prog. Part. Nucl. Phys. 135 (2024). 104094, arXiv:2208.11707 [hep-ph].
[46] J.R. Espinosa, G.F. Giudice, A. Riotto, 2008, arxiv:0710.2484 [hep-ph].

[47] B, D. Micco, Rev. Phys. 5 (2020) 100045. arXiv:1910.00012 [hep-ph].
[48] J.D. de Blas et al., JHEP 01 (2020) 139, arXiv:1905.03764 [hep-ph].
[49] G. Aad et al., ATLAS Collaboration 2022, arXiv:2211.01216 [hep-ex]
[50] A. Tumasyan et al., CMS Collaboration Nature 607 (2022) 60,
[51] E. Bagnaschi et al., Eur. Phys. J. C 79 (2019) 617, arXiv:1905.00892 [hep-ph].
[52] E. Mambrini, N. Nagata, K. A. Olive J. Zheng, Phys. Rev. D 93 (2016) 111703, arXiv:1602.05583 [hep-ph].
[53] A. Held, J. Kwapisz, L. Sartore JHEP 08 (2022) 122, arXiv:2206.04065 [hep-ph].
[54] N. Haba, K. Nagano, Y. Shimizu, T. Yamada, PTEP 2024 (2024) 053B05.
[55] J. Braathen, S. Kanemura, Eur. Phys. J. C 80 (2020) 227, arXiv:1911.11507 [hep-ph].